\documentclass[twocolumn,showpacs,preprintnumbers,amsmath,amssymb]{revtex4}
\usepackage{graphicx}
\usepackage{dcolumn}
\usepackage{bm}
\begin{document}
{\bf Comment on: "Microwave response of a two-dimensional electron
stripe", cond-mat/0407364} \vspace{.15in}

Recently, S.A.Mikhailov et al presented a theoretical
study(cond-mat/0407364) concerning the electromagnetic response of
a finite-width two-dimensional electron stripe. Surprisingly, in
the above manuscript the magnetic field dependence of the
magnetoplasmon spectrum( namely Fig.4 ) reproduces that reported
for the first time in our paper( cond-mat/0405176, JETP Lett., 80,
August 2004, in press). We believe that the above curious
coincidence may result from the authors not to be informed enough.

\noindent M.~V. Cheremisin\\
{\small\indent A.F.Ioffe Physical-Technical Institute\\
\indent 194021, St.Petersburg, Russia \\

\end{document}